\documentclass[aps,prl,twocolumn,groupedaddress,showpacs]{revtex4}
\usepackage{graphicx}
\usepackage{amssymb}

\begin{document}


\title{Low-speed impact craters in loose granular media}


\author{J.S. Uehara, M.A. Ambroso, R.P. Ojha, and D.J. Durian}
\affiliation{UCLA Department of Physics \& Astronomy, Los Angeles, CA 90095-1547}


\date{\today}

\begin{abstract}
	We report on craters formed by balls dropped into dry,
	non-cohesive, granular media.  By explicit variation of ball
	density $\rho_{b}$, diameter $D_{b}$, and drop height $H$, the
	crater diameter is confirmed to scale as the $1/4$ power of
	the energy of the ball at impact:
	$D_{c}\sim(\rho_{b}{D_{b}}^{3}H)^{1/4}$.  Against expectation,
	a different scaling law is discovered for the crater depth:
	$d\sim({\rho_{b}}^{3/2}{D_{b}}^{2}H)^{1/3}$.  The scaling with
	properties of the medium is also established.  The crater
	depth has significance for granular mechanics in that it
	relates to the stopping force on the ball.
\end{abstract}



\maketitle



Sand is fragile: If you set down a ball, no matter how gingerly or how
roughly, the sand can barely support the ball's weight; one slight tap
and the ball digs in deeper.  At the same time, sand can be very
strong: If you drop the ball from some height, the sand can stop it
quickly, forming a shallow crater in the process; a higher drop height
will lead to only a slightly deeper crater.  Similarly intriguing
combinations of toughness and fragility have led to tremendous recent
research activity into the physics of granular media in general
\cite{jnb,duran}.  At the scale of grain-grain interactions, there are
only a few possible forces.  There can be normal forces perpendicular
to the contact plane, there can be static and sliding friction
parallel to the contact plane, and there can be inelastic collisions. 
Since granular packings are random, the normal forces are random too
but can be correlated over long distances in so-called ``force
chains.''  Dissipation can be either through inelastic collisions or
sliding friction.  But which of these is responsible for taking up the
energy of an impacting ball?

Geophysicists have long been interested in the craters formed by
impact or explosion \cite{roddy,melosh,holsapple}.  Since meteorites
generally strike at non-normal angles but nevertheless produce
circular craters, it is believed that impacts can be likened to
explosions.  Extensive data for buried explosives and high-speed
(km/s) ballistics impact indicate that the crater diameter $D_{c}$
often scales as a power of energy.  For example, the exponent is $1/3$
when the energy is dissipated by plastic flow of the medium throughout
a volume $\sim{D_{c}}^{3}$; it is $1/4$ when the energy goes into
lifting a volume $\sim{D_{c}}^{3}$ by a distance $\sim D_{c}$ against
the force of gravity.  This ``gravity-limited'' regime was recently
observed in low-speed laboratory impact experiments~\cite{amato}. 
There, steel balls of various diameters $D_{b}$ were dropped into sand
from various heights $H$; the resulting crater diameters scaled as
$D_{c}\sim({D_{b}}^{3}H)^{1/4}$.  For high-speed impacts in loose
sand, however, ballistics data support $D_{c}\sim {\rho_{b}}^{1/3}
{D_{b}}^{5/6}{H}^{1/6}$ \cite{holsapple,crater}.

Here we focus on the {\it depth} of craters formed by projectile
impact.  By contrast, the diameter has traditionally been the primary
observable.  Prior literature assumes that depth is proportional to
diameter, as in explaining the scaling with energy, but this has not
been checked in laboratory experiments.  There are few actual
observations, but ``simple craters'' formed by meteorite impact are
thought to be parabolic with a depth equal to $1/5$ of the
diameter~\cite{melosh} (``complex craters'' are larger and exhibit
flat floors or central peaks).  The crater depth is also a key
physical quantity because it relates to the force exerted by the
medium onto the projectile.  Specifically, if a ball of mass $m$ is
dropped from rest and forms a crater of depth $d$, then the average
stopping force satisfies
\begin{equation}
    \langle F\rangle d = mgH.  
\label{Econ}
\end{equation}
Note that $H$ is the total drop distance, equal to the sum of initial
height above the medium plus the depth of the crater.  This simple
experiment therefore gives a direct signature of the dissipation
mechanics.  We find that the depth $d$ can be much less than the drop
height $H$; therefore, to a suprising extent, {\it the dissipation
force can far exceed the ball weight}.  We also find that crater depth
and diameter do not exhibit the same scaling with ball density, ball
diameter, and total drop distance.  This contradicts common assumption
\cite{roddy,melosh,holsapple}.  Our observations could lead to new
insight both into fundamental granular forces, and perhaps as well as
into geophysical impact and explosion cratering.  Analogy may also be
found for impact on metals~\cite{utah,zukas}.

Our experimental procedures are as follows.  First, either a 1~L or
2~L beaker is filled half way with a granular medium (results are
independent of beaker size).  The container is then horizontally
swirled and lightly tapped in order to level the surface without
noticeable compaction.  Next a ball is fixed in the jaws of a wrench
mounted to a ringstand.  The ball is released with zero translational
and rotational speed directly above the center of the beaker.  It
splashes into the granular medium and comes to rest with some of its
lower portion buried and its top fully exposed.  This produces a
crater with a circular rim that extends above the original horizontal
surface of the medium.  As is standard, we define the crater diameter
by the location of maximum rim height.  But now, we also measure the
crater depth.  Useful definitions might include the depth above which
the ball is exposed, or the depth below which grains are not
disturbed.  Rather, to make contact with Eq.~(\ref{Econ}) and the
average stopping force on the ball, we define the crater depth as the
distance of the bottom of the ball below the initial horizontal
surface of the medium.  This is determined to $\pm0.5$~mm by measuring
the location of the top exposed surface of the ball.  Before
proceeding, two bounds on the measured quantities should be noted. 
First, it is not possible to measure the crater diameter if it is less
than the ball diameter.  But even if the diameter is less than this
minimum, the depth is still easily measured.  Second, a ball dropped
from infinitesimally above the surface of the medium can still
penetrate to a non-zero depth, more so for denser balls.  This defines
a minimum crater depth, and corresponds to a minimum in total drop
distance $H$.  Our dynamic range for $H$ is $1-1000$~mm at best, but
is more typically a factor of one hundred.

\begin{figure}
\includegraphics[width=3.00in]{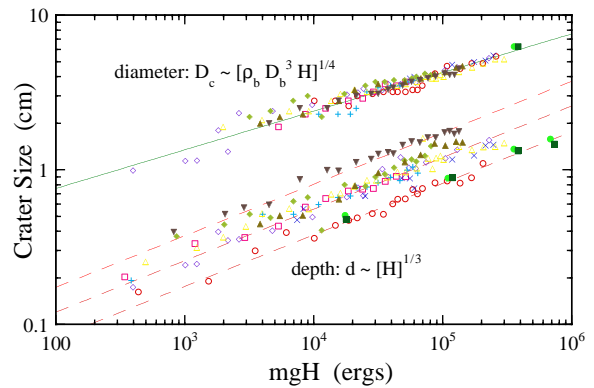}
\caption{The diameter and depth of craters formed by balls dropped
into 0.2~mm diameter glass beads, as a function of total energy loss,
$E=mgH$.  Each data point represents the experimental
result for a single drop height $H$.  Plotting vs $mgH$ collapses the
diameter data, as seen earlier in Ref.~\protect\cite{amato}, onto a
$1/4$ power law.  By contrast, the depth data does not collapse, but
follows a $1/3$ power law.
Symbol, ball type, ball density, and ball diameter are as follows:
$\circ$ for hollow plastic,     $0.26$~g/cm$^3$, $2.54$~cm; 
$\square$ for wood, 	        $0.83$~g/cm$^3$, $1.59$~cm; 
$\lozenge$ for nylon, 	        $1.10$~g/cm$^3$, $1.59$~cm; 
$\times$ for nylon, 	        $1.10$~g/cm$^3$, $2.54$~cm; 
$+$ for silicon rubber,         $1.10$~g/cm$^3$, $1.52$~cm; 
$\triangle$ for acrylic, 	$1.20$~g/cm$^3$, $1.59$~cm; 
$\bullet$ for live ball, 	$1.20$~g/cm$^3$, $3.82$~cm; 
$\blacksquare$ for dead ball, 	$1.30$~g/cm$^3$, $3.82$~cm; 
$\blacklozenge$ for delrin, 	$1.40$~g/cm$^3$, $1.59$~cm; 
$\blacktriangle$ for teflon, 	$2.20$~g/cm$^3$, $1.59$~cm; 
$\blacktriangledown$ for ceramic, 	$3.90$~g/cm$^3$, $1.27$~cm.
NB: Data for three even denser balls are omitted:
stainless steel,                $7.90$~g/cm$^3$, $2.54$~cm;
lead,                           $11.3$~g/cm$^3$, $1.13$~cm; and
tungsten carbide,               $16.4$~g/cm$^3$, $1.91$~cm.
These produce craters that also scale as $D_{c}\sim H^{1/4}$ 
and $d\sim H^{1/3}$; however, they do not exhibit the same $D_{c}\sim 
E^{1/4}$ collapse as the less dense balls.
\label{raw}}
\end{figure}

In Fig.~\ref{raw} we show crater size vs loss in ball energy, $E=mgH$,
for a variety of balls dropped into dry, monodisperse, 0.2~mm diameter
glass beads.  When so plotted, the diameter data collapse onto a
power-law curve, $D_{c}=(0.24~{\rm cm/erg^{1/4}})E^{1/4}$.  The crater
diameter scaling is thus $D_{c}\sim(\rho_{b}{D_{b}}^{3}H)^{1/4}$. 
This agrees with the gravity-limited scaling observation of
Ref.~\cite{amato}.  Our data also extend that work, both in dynamic
range and in explicit variation of the ball density.  However, the
most striking feature of Fig.~\ref{raw} is that the crater depth data
do not show a similar collapse!  Apparently, the crater depth does not
scale with the ball energy.  Furthermore, it appears to follow a
different power-law with total drop distance, $d\sim H^{1/3}$.  The
weighted average of the exponents for power-law fits vs $H$ are
$0.231\pm0.005$ for diameter and $0.318\pm0.005$ for depth; for
individual data sets, the average uncertainty is $0.03$.  Against
natural expectation and contrary to prior
assumption~\cite{roddy,melosh,holsapple,amato}, the crater depth and
diameter are separate length scales set by different physics.

Before further analyzing the data in Fig.~\ref{raw}, two specific
comparisons should be noted.  First, the smaller nylon ball and the
silicon rubber ball have nearly the same density and diameter, but
have very different surface properties: nylon is slick and silicon
rubber is tacky.  But as seen in Fig.~\ref{raw}, the crater depth data
for these two balls ($\lozenge$ and $+$, respectively) are
indistinguishable.  Therefore, friction between the ball and grains
cannot be the dissipation mechanism responsible for stopping the ball. 
Second, the ``live'' and ``dead'' balls have nearly the same density
and diameter, but have very different restitution coefficients.  But
as seen in Fig.~\ref{raw}, the depth data for these two balls 
($\bullet$ and $\blacksquare$, repectively) are indistinguishable. 
Therefore, no significant energy is transfered to internal degrees of
freedom of the ball.

\begin{figure}
\includegraphics[width=3.00in]{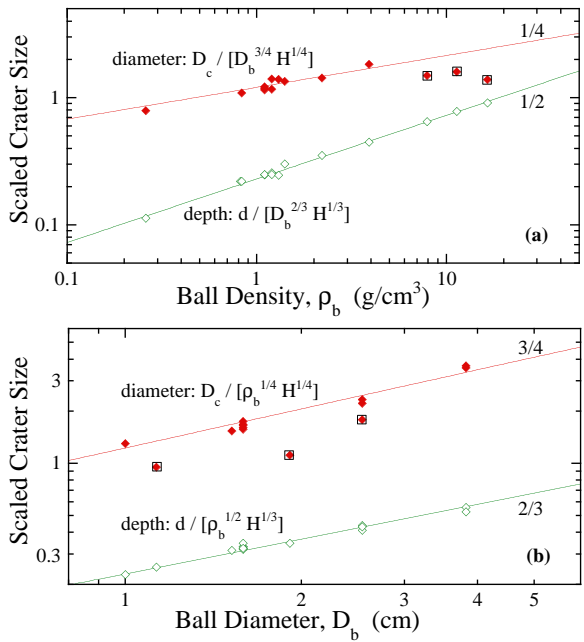}
\caption{Scaling of crater size with ball density and diameter.
Each point represents the result of a $D_{c}\propto H^{1/4}$ or $d\propto
H^{1/3}$ power-law fit to the size vs $H$ data sets of
Fig.~\protect\ref{raw}, divided by the labeled combination of ball diameter
or density and drop height; error bars are comparable to symbol size.
All data are for 0.2~mm diameter glass beads.  Note that
the three densest balls do not obey the $D_{c}\sim{\rho_{b}}^{1/4}$
scaling; data for these balls are boxed in both (a) and (b).}
\label{scale}
\end{figure}

We now explore the full form of the scaling laws for crater size.  If
the ball diameter and drop height are the only relevant length scales,
then the simplest dimensionally-correct laws consistent with the
$H$-dependence of Fig.~\ref{raw} would be $D_{c} \propto
{D_{b}}^{3/4}H^{1/4}$ and $d \propto {D_{b}}^{2/3}H^{1/3}$.  To test
this, we deduce the dimensionless proportionality constants for all
the size vs $H$ data.  The results are plotted in Fig.~\ref{scale}(a)
as a function of ball density.  For the crater diameter, we find
$D_{c}\sim{\rho_{b}}^{1/4}$ as expected from the above $E^{1/4}$
collapse.  However, this trend is violated by the three densest balls
(stainless steel, lead, and tungsten carbide), which were omitted from
Fig.~\ref{raw}.  For these three, the crater diameter is $D_{c}
\approx 1.5 {D_{b}}^{3/4}H^{1/4}$ with no apparent dependence on ball
density.  For the crater depth, we have no {\it a priori} expectation
other than that denser balls should penetrate deeper.  The data in
Fig.~\ref{scale}(a) are consistent with a simple power-law form,
$d\sim {\rho_{b}}^{1/2}$; the uncertainty in the exponent is
conservatively $\pm0.05$.  Note that this holds for {\it all} balls,
even the three densest where $D_{c}\sim{\rho_{b}}^{1/4}$ fails.  As a
final consistency check, the explicit dependence of crater size on
ball diameter is shown in Fig.~\ref{scale}(b).  Though the dynamic
range is only a factor of four, the data are consistent with
expectation, $D_{c}\sim{D_{b}}^{3/4}$ and $d\sim{D_{b}}^{2/3}$.

\begin{figure}
\includegraphics[width=3.00in]{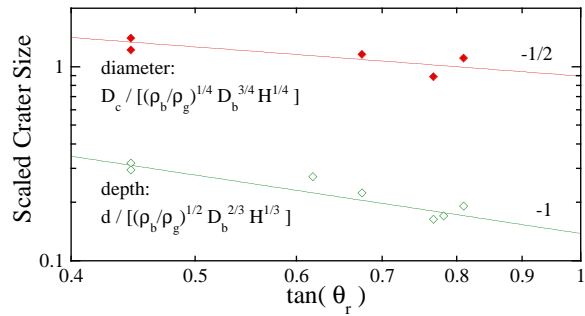}
\caption{Scaling of crater size with grain-grain friction coefficient,
$\mu=\tan\theta_{r}$, for 1-inch diameter nylon ball dropped into the
media specified in Table~\protect\ref{grainspecs}.  Each point
represents the result of a $D_{c}\propto H^{1/4}$ or $d\propto
H^{1/3}$ power-law fit to the size vs $H$ data sets, divided by the
labeled combination of ball and grain densities, ball diameter, and
drop height; error bars are comparable to symbol size.}
\label{grains}
\end{figure}

Next we vary the properties of the granular medium, studying craters
formed by a 1-inch diameter nylon ball dropped into the media listed
in Table~\ref{grainspecs}.  Besides grain size, the other
specifications are the (bulk) grain density $\rho_{g}$, and the angle
of repose $\theta_{r}$.  Since identical cratering is found in 0.2 and
1~mm diameter glass beads, the grain size is not an important length
scale (as suspected already based on the ball diameter and drop height
scaling).  Unless the ambient air plays a crucial role, the dependence
on $\rho_{g}$ must be the reciprocal of the dependence on $\rho_{b}$. 
We have no such guess for behavior vs $\theta_{r}$.  But since
$\tan\theta_{r}$ is roughly the coefficient of friction between
grains, it must play a role.  To test all this we divide out the
expected $\rho_{g}$ dependence and plot the results vs
$\tan\theta_{r}$ in Fig.~\ref{grains}.  The dynamic range is only a
factor of two for both $\rho_{g}$ and $\mu=\tan\theta_{r}$, so
power-law fits give exponents to within only $\pm 0.3$.  Nevertheless,
the data are consistent with $D\sim(\rho_{g}\mu^{2})^{-1/4}$ for the
crater diameter and $d\sim(\rho_{g}\mu^{2})^{-1/2}$ for the crater
depth.  Comfortingly, the same combination of material properties
appears in both results, and also leads to a stopping force that is
proportional to $\mu$ (next).

\begin{table}[ht]
\begin{ruledtabular}
\begin{tabular}{lccc}
Material & Grain Size (mm) & $\rho_{g}$ (g/cm$^{3}$) & $\theta_{r}$ \\
\colrule
sprinkles & $2\times7$ & 0.76 & $39^{\circ}$ \\
popcorn & $4\times6\times7$ & 0.87 & $34^{\circ}$ \\
rice & $2\times7$ & 0.88 & $32^{\circ}$ \\
salt & 0.5 & 1.30 & $38^{\circ}$ \\
glass beads & 0.2 or 1.0 & 1.51 & $24^{\circ}$ \\
beach sand & $0.5\pm0.4$ & 1.59 & $38^{\circ}$ \\
\end{tabular}
\end{ruledtabular}
\caption{Specifications of the granular materials.  The density refers
to the bulk material, not the individual grains.  The angle of repose,
$\theta_{r}$, was measured by the draining method
\protect\cite{brown}, which seems most appropriate for craters; it
gives the grain-grain friction coefficient as $\mu=\tan\theta_{r}$. 
The beach sand is the only significantly polydisperse material.
\label{grainspecs}}
\end{table}

The final crater diameter and depth laws established by
Figs.~\ref{raw}-\ref{grains} are, respectively:
\begin{eqnarray}
  D_{c} &=& 
  0.92[\rho_{b}/(\rho_{g}\mu^{2})]^{1/4}{D_{b}}^{3/4}H^{1/4},\label{diam} \\
  d &=& 0.16[\rho_{b}/(\rho_{g}\mu^{2})]^{1/2}{D_{b}}^{2/3}H^{1/3}.\label{depth}
\end{eqnarray}
Whereas the depth law holds for all our observations, the diameter law
fails for dense balls, $\rho_{b}>4$~g/cm$^{3}$ (at least for glass
beads - the materials dependence of this breakdown has not been
investigated).  For a fixed granular material, the diameter data
collapse onto a $1/4$ power-law when plotted vs $\rho_{b}{D_{b}}^{3}H$
(as in Fig.~\ref{raw}); therefore, the diameter scales with ball
energy at impact.  By contrast, the depth data collapse onto a $1/3$
power-law when plotted vs ${\rho_{b}}^{3/2}{D_{b}}^{2}H$ (not shown -
but since the scatter for depth and diameter is the same in
Figs.~\ref{scale}-\ref{grains}, the quality of collapse is also
comparable); therefore, the depth scales as neither ball energy nor
ball momentum at impact.  The diameter and depth are separate lengths
set by separate physics.

The crater depth $d$ may be the more fundamental length scale, in that
it naturally relates to the underlying granular mechanics.  Energy
conservation, Eq.~(\ref{Econ}), and the depth law, Eq.~(\ref{depth}),
give the {\it average} stopping force acting on the ball as
\begin{equation}
   {\langle F\rangle \over mg}
   = 6.3\mu \left( \rho_{g}\over\rho_{b}\right)^{1/2}
            \left( H \over D_{b} \right)^{2/3}.
\label{AvgForce}
\end{equation}
Evidently, it can be much greater than the ball weight, $mg$, which
the granular medium can barely support in a static situation.  These
results help constrain the form of the {\it instantaneous} stopping
force on the ball throughout the impact process.  If it varies only
with ball speed, independent of depth, then it must scale as $F\sim
v^{4/3}$.  Alternatively, if the stopping force varies only with ball
depth, independent of speed, then it must scale as $F\sim z^{2}$. 
Solution of $F=ma$ gives $d\sim H^{1/3}$ for both cases.

%
%

We now discuss these possible force laws in the context of the current
understanding of granular mechanics.  First, the kinetic theory of
granular hydrodynamics gives a rate-dependent stopping force
\cite{bagnold,savage}.  In Ref.~\cite{siggi} it was used to analyze
the peak height of the granular jet formed when a heavy ball strikes
the medium and becomes deeply submerged.  There, the viscosity of the
medium is proportional to the shear rate, which scales as ball speed
divided by diameter.  This gives a stopping force of $F\propto
\rho_{g}{D_{g}}^{2}v^{2}$, where $D_{g}$ is the grain diameter and the
numerical constant is set by the grain-grain restitution coefficient. 
This cannot account for our observations, since the dependencies on
grain density, grain size, ball size, and ball speed are all wrong. 
Furthermore, a $F\sim v^{2}$ drag force is not even strong enough to
bring an object to rest!  A modification, whereby the viscosity
increases with packing fraction~\cite{penn2}, would be required.

Second, the lateral drag force on an object slowly pulled through a
granular medium at constant depth is rate-independent, as in plowing a
field.  Recently it was found to scale as the product of the object's
cross section and the hydrostatic pressure at that
depth~\cite{schiffer01}.  If this applies to our work, where the ball
moves down rather than sideways, then the instantaneous drag force is
proportional to the weight of displaced grains: $F= \eta \rho_{g}g \pi
( D_{b}z^{2}/2-z^{3}/3 )$.  Here $z$ is the depth of the bottom of the
ball, the volume of displaced grains equals the submerged volume of
the ball, and $\eta$ is a materials parameter like $\mu$.  For $z \ll
D_{b}$, the leading term is $F\sim z^{2}$ and the final crater depth
is $d = [ (\rho_{b}/\rho_{g}){D_{b}}^{2}H/\eta]^{1/3}$.  This is quite
similar to our observation, Eq.~(\ref{depth}); however the ball- and
grain density dependencies are incorrect.  The parameter range in
Fig.~\ref{scale}(a) is great enough to easily rule out $d\sim
{\rho_{b}}^{1/3}$ in favor of $d\sim {\rho_{b}}^{1/2}$.

Altogether, it appears that our observation for the scaling of crater
depth, Eq.~(\ref{depth}), cannot be explained using prior work. 
Friction at the ball surface, collisional granular hydrodynamics, and
the plowing of hydrostatic grains, are all ruled out.  One possible
scenario is that as the ball crashes into the medium, it jams
together the grains underneath.  The normal force between these grains
thus becomes much greater than the hydrostatic pressure.  As the ball
moves, the grain contacts slide so that each dissipates a total amount
of energy given by the normal force times grain size.  New contacts
are formed as the old ones break.  This loading and breaking of force
chains gives rise to the dissipation force that ultimately stops the
ball.  Another possibility is that dissipation is due to sliding
friction between force chains and the surrounding unloaded grains. 
Further work is needed to model either effect.  Perhaps the scaling of
crater diameter, Eq.~(\ref{diam}), could then be explained as a
consequence of the stopping force between ball and grains.

We thank R.P. Behringer, S.R. Nagel, and J.A. Rudnick for helpful
suggestions.  This material is based upon work supported by the
National Science Foundation under Grant No.~0070329.

{\it Note added in proof.} A similar experiment has been reported, in
which a steel ball was dropped into glass beads of several different
sizes~\cite{deBruyn}.  There, the impact energy was greater and the
granular packing may have been looser, so that the ball always became
submerged.  The crater diameter scaled as $H^{1/4}$; the crater depth
(measured to the bottom of the crater, not the bottom of the ball) was
about $1/8$ the diameter but increased with $H$.  The crater
morphology and the analogy with planetary craters were emphasized.

\bibliography{craterrefs}

\begin{thebibliography}{16}
\expandafter\ifx\csname natexlab\endcsname\relax\def\natexlab#1{#1}\fi
\expandafter\ifx\csname bibnamefont\endcsname\relax
  \def\bibnamefont#1{#1}\fi
\expandafter\ifx\csname bibfnamefont\endcsname\relax
  \def\bibfnamefont#1{#1}\fi
\expandafter\ifx\csname citenamefont\endcsname\relax
  \def\citenamefont#1{#1}\fi
\expandafter\ifx\csname url\endcsname\relax
  \def\url#1{\texttt{#1}}\fi
\expandafter\ifx\csname urlprefix\endcsname\relax\def\urlprefix{URL }\fi
\providecommand{\bibinfo}[2]{#2}
\providecommand{\eprint}[2][]{\url{#2}}

\bibitem[{\citenamefont{Jaeger et~al.}(1996)\citenamefont{Jaeger, Nagel, and
  Behringer}}]{jnb}
\bibinfo{author}{\bibfnamefont{H.~M.} \bibnamefont{Jaeger}},
  \bibinfo{author}{\bibfnamefont{S.~R.} \bibnamefont{Nagel}}, \bibnamefont{and}
  \bibinfo{author}{\bibfnamefont{R.~P.} \bibnamefont{Behringer}},
  \bibinfo{journal}{Rev. Mod. Phys.} \textbf{\bibinfo{volume}{68}},
  \bibinfo{pages}{1259} (\bibinfo{year}{1996}).

\bibitem[{\citenamefont{Duran}(2000)}]{duran}
\bibinfo{author}{\bibfnamefont{J.}~\bibnamefont{Duran}},
  \emph{\bibinfo{title}{Sands, powders, and grains: An introduction to the
  physics of granular materials}} (\bibinfo{publisher}{Springer},
  \bibinfo{address}{NY}, \bibinfo{year}{2000}).

\bibitem[{\citenamefont{Roddy et~al.}(1978)\citenamefont{Roddy, Pepin, and
  Merril}}]{roddy}
\bibinfo{editor}{\bibfnamefont{D.~J.} \bibnamefont{Roddy}},
  \bibinfo{editor}{\bibfnamefont{R.~O.} \bibnamefont{Pepin}}, \bibnamefont{and}
  \bibinfo{editor}{\bibfnamefont{R.~B.} \bibnamefont{Merril}}, eds.,
  \emph{\bibinfo{title}{Impact and explosion cratering}}
  (\bibinfo{publisher}{Pergamon Press}, \bibinfo{address}{NY},
  \bibinfo{year}{1978}).

\bibitem[{\citenamefont{Melosh}(1989)}]{melosh}
\bibinfo{author}{\bibfnamefont{H.~J.} \bibnamefont{Melosh}},
  \emph{\bibinfo{title}{Impact cratering: A geologic process}}
  (\bibinfo{publisher}{Oxford University Press}, \bibinfo{address}{NY},
  \bibinfo{year}{1989}).

\bibitem[{\citenamefont{Holsapple}(1993)}]{holsapple}
\bibinfo{author}{\bibfnamefont{K.}~\bibnamefont{Holsapple}},
  \bibinfo{journal}{Ann. Rev. Earth Plan. Sci.} \textbf{\bibinfo{volume}{21}},
  \bibinfo{pages}{333} (\bibinfo{year}{1993}).

\bibitem[{\citenamefont{Amato and Williams}(1998)}]{amato}
\bibinfo{author}{\bibfnamefont{J.}~\bibnamefont{Amato}} \bibnamefont{and}
  \bibinfo{author}{\bibfnamefont{R.}~\bibnamefont{Williams}},
  \bibinfo{journal}{Am. J. Phys.} \textbf{\bibinfo{volume}{66}},
  \bibinfo{pages}{141} (\bibinfo{year}{1998}).

\bibitem[{\citenamefont{Melosh and Beyer}(1999)}]{crater}
\bibinfo{author}{\bibfnamefont{H.~J.} \bibnamefont{Melosh}} \bibnamefont{and}
  \bibinfo{author}{\bibfnamefont{R.~A.} \bibnamefont{Beyer}}
  (\bibinfo{year}{1999}), \bibinfo{note}{computer code CRATER, University of
  Arizona, Tuscon AZ (available at
  http://www.lpl.arizona.edu/tekton/crater.html)}.

\bibitem[{\citenamefont{van Valkenburg et~al.}(1956)\citenamefont{van
  Valkenburg, Clay, and Huth}}]{utah}
\bibinfo{author}{\bibfnamefont{M.~E.} \bibnamefont{van Valkenburg}},
  \bibinfo{author}{\bibfnamefont{W.~G.} \bibnamefont{Clay}}, \bibnamefont{and}
  \bibinfo{author}{\bibfnamefont{J.~H.} \bibnamefont{Huth}},
  \bibinfo{journal}{J. Appl. Phys.} \textbf{\bibinfo{volume}{27}},
  \bibinfo{pages}{1123} (\bibinfo{year}{1956}).

\bibitem[{\citenamefont{Zukas}(1990)}]{zukas}
\bibinfo{editor}{\bibfnamefont{J.~A.} \bibnamefont{Zukas}}, ed.,
  \emph{\bibinfo{title}{High velocity impact dynamics}}
  (\bibinfo{publisher}{Wiley}, \bibinfo{address}{NY}, \bibinfo{year}{1990}).

\bibitem[{\citenamefont{Brown and Richards}(1970)}]{brown}
\bibinfo{author}{\bibfnamefont{R.~L.} \bibnamefont{Brown}} \bibnamefont{and}
  \bibinfo{author}{\bibfnamefont{J.~C.} \bibnamefont{Richards}},
  \emph{\bibinfo{title}{Principles of Powder Mechanics}}
  (\bibinfo{publisher}{Pergamon Press}, \bibinfo{address}{Oxford},
  \bibinfo{year}{1970}).

\bibitem[{\citenamefont{Bagnold}(1954)}]{bagnold}
\bibinfo{author}{\bibfnamefont{R.}~\bibnamefont{Bagnold}},
  \bibinfo{journal}{Proc. R. Soc. London Ser. A}
  \textbf{\bibinfo{volume}{225}}, \bibinfo{pages}{49} (\bibinfo{year}{1954}).

\bibitem[{\citenamefont{Savage and Sayed}(1984)}]{savage}
\bibinfo{author}{\bibfnamefont{S.}~\bibnamefont{Savage}} \bibnamefont{and}
  \bibinfo{author}{\bibfnamefont{M.}~\bibnamefont{Sayed}}, \bibinfo{journal}{J.
  Fluid Mech.} \textbf{\bibinfo{volume}{142}}, \bibinfo{pages}{391}
  (\bibinfo{year}{1984}).

\bibitem[{\citenamefont{Thoroddsen and Shen}(2001)}]{siggi}
\bibinfo{author}{\bibfnamefont{S.}~\bibnamefont{Thoroddsen}} \bibnamefont{and}
  \bibinfo{author}{\bibfnamefont{A.}~\bibnamefont{Shen}},
  \bibinfo{journal}{Phys. Fluids} \textbf{\bibinfo{volume}{13}},
  \bibinfo{pages}{4} (\bibinfo{year}{2001}).

\bibitem[{\citenamefont{Bocquet et~al.}(2002)\citenamefont{Bocquet, Losert,
  Schalk, Lubensky, and Gollub}}]{penn2}
\bibinfo{author}{\bibfnamefont{L.}~\bibnamefont{Bocquet}},
  \bibinfo{author}{\bibfnamefont{W.}~\bibnamefont{Losert}},
  \bibinfo{author}{\bibfnamefont{D.}~\bibnamefont{Schalk}},
  \bibinfo{author}{\bibfnamefont{T.~C.} \bibnamefont{Lubensky}},
  \bibnamefont{and} \bibinfo{author}{\bibfnamefont{J.~P.}
  \bibnamefont{Gollub}}, \bibinfo{journal}{Phys. Rev. E}
  \textbf{\bibinfo{volume}{65}}, \bibinfo{pages}{011307}
  (\bibinfo{year}{2002}).

\bibitem[{\citenamefont{Albert et~al.}(2001)\citenamefont{Albert, Sample,
  Morss, Rajagopalan, Barabasi, and Schiffer}}]{schiffer01}
\bibinfo{author}{\bibfnamefont{I.}~\bibnamefont{Albert}},
  \bibinfo{author}{\bibfnamefont{J.~G.} \bibnamefont{Sample}},
  \bibinfo{author}{\bibfnamefont{A.~J.} \bibnamefont{Morss}},
  \bibinfo{author}{\bibfnamefont{S.}~\bibnamefont{Rajagopalan}},
  \bibinfo{author}{\bibfnamefont{A.~L.} \bibnamefont{Barabasi}},
  \bibnamefont{and} \bibinfo{author}{\bibfnamefont{P.}~\bibnamefont{Schiffer}},
  \bibinfo{journal}{Phys. Rev. E} \textbf{\bibinfo{volume}{64}},
  \bibinfo{pages}{061303/1} (\bibinfo{year}{2001}), \bibinfo{note}{and
  references therein}.

\bibitem[{\citenamefont{Walsh et~al.}(2003)\citenamefont{Walsh, Holloway,
  Habdas, and de~Bruyn}}]{deBruyn}
\bibinfo{author}{\bibfnamefont{A.~M.} \bibnamefont{Walsh}},
  \bibinfo{author}{\bibfnamefont{K.~E.} \bibnamefont{Holloway}},
  \bibinfo{author}{\bibfnamefont{P.}~\bibnamefont{Habdas}}, \bibnamefont{and}
  \bibinfo{author}{\bibfnamefont{J.~R.} \bibnamefont{de~Bruyn}},
  \bibinfo{journal}{preprint}  (\bibinfo{year}{2003}).

\end{thebibliography}

\end{document}